\documentclass[sigconf,nonacm=true]{acmart}

\usepackage[utf8]{inputenc}
\DeclareUnicodeCharacter{2212}{-}
\usepackage{pgfplots}
\usepgfplotslibrary{groupplots}
\usepgfplotslibrary{dateplot}
\usetikzlibrary{patterns}
\usetikzlibrary{matrix}
\usepackage{tikz}
\usepackage{tikzscale}
\usepackage{graphicx}
\usepackage[nohyperlinks, printonlyused, nolist]{acronym}
\usepackage{mdwlist}
\usepackage{subcaption}
\usepackage[disable]{todonotes}
\usepackage{listings}
\usepackage{mdwlist}
\usepackage{balance}
\usepackage{xspace}
\usepackage{titlesec}
\usepackage{caption}
\usepackage{subcaption}

\titlespacing{\paragraph}{5pt}{2pt}{2pt}
\titlespacing{\subsection}{1pt}{5pt}{2pt}
\titlespacing{\subsubsection}{1pt}{5pt}{2pt}

\newcommand{\sysname}{AnyDB\xspace}

\newcommand{\mimic}{mimic\xspace}
\newcommand{\mimics}{mimics\xspace}
\newcommand{\mimicking}{mimicking\xspace}

\newcommand{\naive}{na\"{\i}ve\xspace}

\def\BibTeX{{\rm B\kern-.05em{\sc i\kern-.025em b}\kern-.08emT\kern-.1667em\lower.7ex\hbox{E}\kern-.125emX}}

\copyrightyear{2021}
\acmYear{2021}
\setcopyright{none}

\copyrightyear{2021} 
\acmYear{2021} 
\setcopyright{none}
\acmConference[CIDR'21]{2021 Conference on Innovative Data Systems Research}{January 10-13, 2021}{Asilomar, California}
\acmBooktitle{CIDR'21: Conference on Innovative Data Systems Research, January 13-16, 2019, Asilomar, California}
\acmPrice{15.00}
\acmDOI{TBD}
\acmISBN{TBD}

\begin{document}

\title{AnyDB: An Architecture-less DBMS for Any Workload}

\author{Tiemo Bang}
\affiliation{TU Darmstadt \& SAP SE}

\author{Norman May}
\affiliation{SAP SE}

\author{Ilia Petrov}
\affiliation{Reutlingen University}

\author{Carsten Binnig}
\affiliation{TU Darmstadt}

\begin{abstract}
In this paper, we propose a radical new approach for scale-out distributed DBMSs.
Instead of hard-baking an architectural model, such as a shared-nothing architecture, into the distributed DBMS design, we aim for a new class of so-called architecture-less DBMSs.
The main idea is that an architecture-less DBMS can \mimic{} any architecture on a per-query basis on-the-fly without any additional overhead for reconfigura\-tion.
Our initial results show that our architecture-less DBMS \sysname{} can provide significant speed-ups across varying workloads compared to a traditional DBMS implementing a static architecture.
\end{abstract}

\maketitle

\pgfplotsset{
    compat=newest,
}
\pgfplotsset{every x tick label/.append style={font=\small, yshift=0.5ex}}
\pgfplotsset{every y tick label/.append style={font=\small, xshift=0.4ex}}
\pgfplotsset{tick label style = {font = \small}, label style = {font = \small}, legend style = {font = \small}, title style = {font = \small}, every axis plot/.append style = {font = \tiny}}
\pgfplotsset{legend cell align={left}}
\pgfplotsset{legend image post style={scale=0.7,mark options={scale=1}}}

\newlength\figH
\newlength\figW
\setlength\figH{3.4cm}
\setlength\figW{0.52\columnwidth}

\section{Introduction}
\label{sec:introduction}

\paragraph{Motivation:} Scale-out distributed architectures are used today by many academic and commercial database systems such as SAP HANA, Amazon Redshift / Aurora, and Snowflake~\cite{RN397,RN398,RN399} to process large data volumes since they allow scaling compute and memory capacities by simply adding or removing processing nodes.
The two predominant architectural models used in academic and commercial distributed databases are the shared-nothing (aggregated) architecture and the shared-disk (disaggregated) architecture.

While the shared-nothing (aggregated) architecture provides high performance in case the data and workload are well partitionable, its performance degrades significantly under skew, since some resources are overloaded while others are idle.
Moreover, dealing with requirements such as elasticity in the shared-nothing architecture is hard, since this always requires repartitioning the data even if compute is the bottleneck.
This renders the shared-nothing architecture less suited for modern environments such as the cloud where elasticity is a key requirement \cite{repart1}.

On the other hand, the shared-disk (disaggregated) architecture tackles the drawbacks of the shared-nothing architecture by separating storage from compute. 
This separation provides many new potentials especially for better skew handling as well as providing elasticity independently for compute and storage. 
Yet, the shared-disk (disaggregated) architecture has other downsides. 
One major downside is that data always needs to be pulled into the compute layer, resulting in higher latencies.
While this additional latency often does not matter for OLAP, it renders this architecture less interesting for OLTP which requires low latency execution to reduce the potential of conflicts and provide high throughput.

Another observation is that these architectural models (shared-nothing or shared-disk) are statically baked into the system design of today's databases \cite{config}.
However, modern workloads are versatile, e.g., HTAP containing a mix of OLTP and OLAP queries. 
Moreover, workloads are often evolving over time or as in the cloud are not even foreseeable for cloud providers, since workloads are user-specific.
Hence, databases following a static architectural model are always a kind-of-a compromise, since they cannot provide optimal performance across a wide-spectrum of workloads.

\begin{figure}
    \centering
    
    \begin{tikzpicture}
        \setlength\figW{1.1\linewidth}
        \setlength\figH{4cm}
\definecolor{color0}{rgb}{0.298039215686275,0.447058823529412,0.690196078431373}
\definecolor{color1}{rgb}{0.866666666666667,0.517647058823529,0.32156862745098}

\begin{axis}[
title style = {text depth=0ex},
axis line style={white!15!black},
height=\figH,
legend cell align={left},
legend entries={DBx1000, AnyDB},
legend style={fill=none, text opacity=1, draw=white!80!black, at={(0.025,0.1)}, anchor=south west, nodes={scale=0.7, transform shape}},
tick pos=left,
title style={yshift=-1.5ex},
title={OLTP Performance},
width=\figW,
x grid style={white!80!black},
xlabel shift=-1ex,
xlabel={Workload Phase},
xmajorticks=true,
xmin=-0.55, xmax=11.55,
xtick style={color=white!15!black},
xticklabel style={align=left},
y grid style={white!80!black},
ylabel shift=-1.4ex,
ytick scale label code/.code={},
ylabel={Throughput (M tx/s)},
ymin=0, ymax=2757745.0955845,
yminorticks=false,
ytick style={color=white!15!black},
]
\path [draw=color0, line width=0.48pt]
(axis cs:0,2464100.2801027)
--(axis cs:0,2473585.7740863);

\path [draw=color0, line width=0.48pt]
(axis cs:1,2459398.72376511)
--(axis cs:1,2469355.41488536);

\path [draw=color0, line width=0.48pt]
(axis cs:2,2458627.03878892)
--(axis cs:2,2469107.83220649);

\path [draw=color0, line width=0.48pt]
(axis cs:3,725725.727201536)
--(axis cs:3,741116.999386462);

\path [draw=color0, line width=0.48pt]
(axis cs:4,728100.359760864)
--(axis cs:4,736059.579628886);

\path [draw=color0, line width=0.48pt]
(axis cs:5,712855.221080549)
--(axis cs:5,738640.219401011);

\path [draw=color0, line width=0.48pt]
(axis cs:6,640556.094924483)
--(axis cs:6,649764.406323123);

\path [draw=color0, line width=0.48pt]
(axis cs:7,640011.984426646)
--(axis cs:7,646238.884883441);

\path [draw=color0, line width=0.48pt]
(axis cs:8,638415.756565759)
--(axis cs:8,651997.971836978);

\path [draw=color0, line width=0.48pt]
(axis cs:9,1354636.40183172)
--(axis cs:9,1362246.73651445);

\path [draw=color0, line width=0.48pt]
(axis cs:10,1352450.5029531)
--(axis cs:10,1359649.57792957);

\path [draw=color0, line width=0.48pt]
(axis cs:11,1358054.45119322)
--(axis cs:11,1364443.7887534);

\path [draw=color1, line width=0.48pt]
(axis cs:0,2375032.47114707)
--(axis cs:0,2390694.81809427);

\path [draw=color1, line width=0.48pt]
(axis cs:1,2372400.09044775)
--(axis cs:1,2379745.68698982);

\path [draw=color1, line width=0.48pt]
(axis cs:2,2376528.21904461)
--(axis cs:2,2388582.87244258);

\path [draw=color1, line width=0.48pt]
(axis cs:3,1659510.02966374)
--(axis cs:3,1688661.48247592);

\path [draw=color1, line width=0.48pt]
(axis cs:4,1655567.46644372)
--(axis cs:4,1706397.28341552);

\path [draw=color1, line width=0.48pt]
(axis cs:5,1650020.36743891)
--(axis cs:5,1688107.38895155);

\path [draw=color1, line width=0.48pt]
(axis cs:6,1359800.72120431)
--(axis cs:6,1409391.12541643);

\path [draw=color1, line width=0.48pt]
(axis cs:7,1360679.59142194)
--(axis cs:7,1421942.88944127);

\path [draw=color1, line width=0.48pt]
(axis cs:8,1362140.33110226)
--(axis cs:8,1395725.76431251);

\path [draw=color1, line width=0.48pt]
(axis cs:9,2176710.08636437)
--(axis cs:9,2194699.32104919);

\path [draw=color1, line width=0.48pt]
(axis cs:10,2181755.95218818)
--(axis cs:10,2196743.05382982);

\path [draw=color1, line width=0.48pt]
(axis cs:11,2177520.30112384)
--(axis cs:11,2198070.1630865);

\addplot [line width=0.48pt, color0, mark=*, mark size=2, mark options={solid,draw=white}]
table {%
0 2468935.80098019
1 2464420.6923849
2 2463741.93052356
3 736083.717521005
4 728692.796478427
5 730132.873875364
6 645838.011103938
7 643150.19856667
8 644152.8010814
9 1359224.86803838
10 1357038.8225483
11 1360826.56605622
};
\addplot [line width=0.48pt, color1, mark=triangle*, mark size=2, mark options={solid,rotate=0,draw=white}]
table {%
0 2383030.44023509
1 2376234.38878982
2 2379494.40948942
3 1679943.46990224
4 1676198.2984492
5 1672149.92496227
6 1363224.29810989
7 1365287.89655214
8 1366056.06806374
9 2190242.40165555
10 2195455.95503706
11 2186878.32004007
};
\addplot [line width=0.48pt, white!50.1960784313725!black, dashed, forget plot]
table {%
2.5 0
2.5 2757745.0955845
};
\addplot [line width=0.48pt, white!50.1960784313725!black, dashed, forget plot]
table {%
5.5 0
5.5 2757745.0955845
};
\addplot [line width=0.48pt, white!50.1960784313725!black, dashed, forget plot]
table {%
8.5 0
8.5 2757745.0955845
};
\addplot [line width=0.48pt, color0, forget plot]
table {%
0 2468935.80098019
1 2464420.6923849
2 2463741.93052356
3 736083.717521005
4 728692.796478427
5 730132.873875364
6 645838.011103938
7 643150.19856667
8 644152.8010814
9 1359224.86803838
10 1357038.8225483
11 1360826.56605622
};
\addplot [line width=0.48pt, color1, forget plot]
table {%
0 2383030.44023509
1 2376234.38878982
2 2379494.40948942
3 1679943.46990224
4 1676198.2984492
5 1672149.92496227
6 1363224.29810989
7 1365287.89655214
8 1366056.06806374
9 2190242.40165555
10 2195455.95503706
11 2186878.32004007
};
\draw (axis cs:1,2537125.48793774) node[
  scale=0.6,
  anchor=base,
  text=white!15!black,
  rotate=0.0,
  align=center
]{OLTP
partitionable};
\draw (axis cs:4,2537125.48793774) node[
  scale=0.6,
  anchor=base,
  text=white!15!black,
  rotate=0.0,
  align=center
]{OLTP
skewed};
\draw (axis cs:7,2537125.48793774) node[
  scale=0.6,
  anchor=base,
  text=white!15!black,
  rotate=0.0,
  align=center
]{HTAP
skewed};
\draw (axis cs:10,2537125.48793774) node[
  scale=0.6,
  anchor=base,
  text=white!15!black,
  rotate=0.0,
  align=center
]{HTAP
partitionable};
\end{axis}
    \end{tikzpicture}
    \vspace{-2ex}
    \caption{Performance of \sysname{} across a workload evolving from partitionable OLTP (phase 0-2), over a skewed OLTP (phase 3-5), to skewed HTAP (phase 6-8), and then to partitionable HTAP (phase 9-11).  
    The y-axis only shows the throughput of the OLTP transactions excluding the OLAP queries in the HTAP phases.}
    \label{fig:headline}
    \vspace{-3.5ex}
\end{figure}
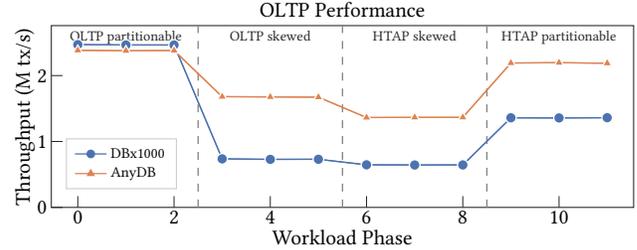

\paragraph{Contributions:} In this paper, we propose a radical new approach for scale-out distributed DBMSs.
Instead of hard-bak\-ing an architectural model statically into the DBMS design, we aim for a new class of so-called \emph{architecture-less DBMSs}.

The main idea of an architecture-less database system is that it is composed of a single generic type of component where multiple instances of those components ``act together'' in an optimal manner on a per-query basis.
To instrument generic components at runtime and coordinate the overall DBMS execution, each component consumes two streams: an event and a data stream. While the event stream encodes the operations to be executed, the data stream shuffles the state required by these events to the executing component.
Through this instrumentation of generic components by event and data streams,
a component can act as a query optimizer at one moment for one query but for the next as a worker executing a filter or join operator.

A key aspect of this execution model is that by simply changing the event and data routing between generic components, an architecture-less DBMS can \mimic{} different distributed architectures and form traditional architectures as well as completely new architectures.
Another important aspect is, since we decide the event and data routing on a per-query basis, an architecture-less DBMS can simultaneously act as a shared-nothing system for one query while acting as a shared-disk (disaggregated) DBMS for another query that runs concurrently with the first query.
This opens up interesting opportunities for running mixed workloads, e.g., HTAP, or adapting the execution to evolving workloads.

Also, an interesting aspect of this execution model is that it cannot only \mimic{} architectures on the macro-level (shared-nothing vs. shared-disk) but also can adapt execution strategies on the micro-level.
For example, query execution in an architecture-less DBMS can \mimic{} various query processing models at runtime (tuple-wise pipelined vs. vectorized vs. materialized) and degrees of parallelism by simply instrumenting the generic components with different event and data streams.
The same holds also for other components such as transaction execution and concurrency control.  

The potential of the architecture-less database system is shown in Figure \ref{fig:headline}. Here, we compared the performance when running an evolving workload in a static shared-nothing architecture (blue line) based on an extended version of DBx1000~\cite{abyss} 
compared to \sysname{} (orange line) our prototypical implementation of an architecture-less database system.
As we see, \sysname{} is either able to meet the performance of DBx1000 if its static architecture is optimal for the workload or outperform DBx1000 where the static architecture is not optimal, e.g., skewed workloads.

\paragraph{Outline:} The remainder of this paper is structured as follows.
First, in Section \ref{sec:overview} we give an overview of how we envision an architecture-less database system.
Second, in Sections \ref{sec:oltp}
and \ref{sec:olap} we then discuss the opportunities that an architecture-less database provides for OLTP, OLAP as well as HTAP and present initial experimental results in each of these sections using our prototypical architecture-less database system \sysname{}.
Finally, we conclude with a discussion of future directions in Section \ref{sec:directions}.
\section{An Architecture-less DBMS}
\label{sec:overview}

In the following, we first give an overview of the general execution model of an architecture-less DBMS.
Afterwards we then present how typical database workloads can be mapped to this execution model and discuss the main challenges of this model.

\subsection{Overview of Execution Model}
\label{sec:overview:execution}

\begin{figure}
\centering
    \includegraphics[width=1.0\linewidth]{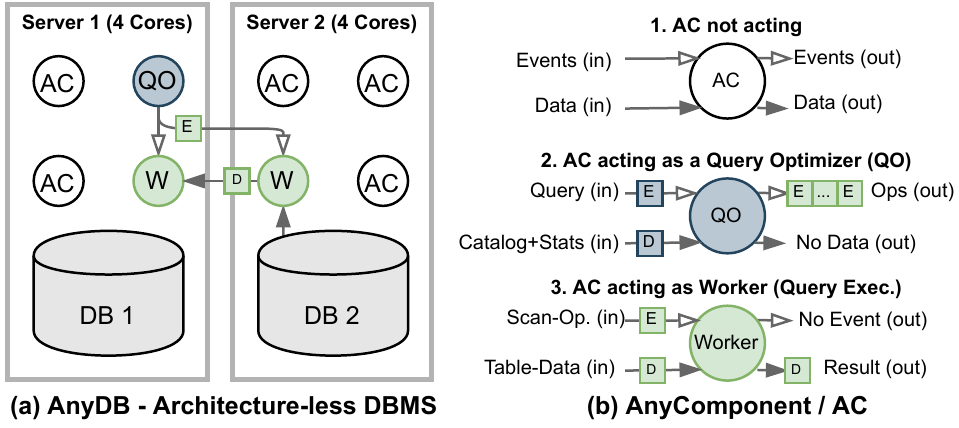}
   \vspace{-5.5ex}
   \caption[]{\sysname{} is an architecture-less DBMS of generic components called AnyComponents
   (ACs), executing arbitrary logic. ACs are instrumented by events and data streams. Depending on the incoming events an AC can act as a Query Optimizer (QO) or a Worker (W) executing a scan or a join operator or any other component, e.g., log writer, etc. An AC can also produce new data and event streams for other ACs. For example, an AC that acts as a scan operator produces a data stream with results of the scan operation.}
   \Description{\sysname{} is an architecture-less database system composed of generic components called AnyComponents
   (ACs), executing arbitrary logic. ACs are instrumented by events and data streams. Depending on the incoming events an AC can act as a Query Optimizer (QO) or a Worker (W) executing a scan or a join operator or any other component, e.g., log writer, etc. An AC can also produce new data and event streams and send these to other ACs. For example, an AC that acts as a scan operator produces a data stream with results of the scan operation.}
	\label{fig:anydb1}
	\vspace{-2.5ex}
\end{figure}

As shown in Figure \ref{fig:anydb1}, the main idea of an architecture-less database system such as \sysname{} is that the DBMS is composed only of generic components, so-called \emph{AnyComponents} (ACs).
These generic ACs can provide any database functionality.
By routing events and their required data to an AC, the AC can act as a query optimizer in one moment and in the next moment as a query executor or any other component (e.g., log writer, etc.).
This gives an architecture-less DBMS the flexibility to shift its architecture just in an instant without any downtime for reconfiguration, as we discuss later.

To execute a complete SQL query (or a transaction composed of a set of database  operations), multiple events and data streams are routed through \sysname{} from one AC to another AC.
For example, as outlined in Figure \ref{fig:anydb1} (a), when executing a query in \sysname{}, calling the query optimizer is one event that can trigger follow-up events for executing the operators, e.g., scans and joins.
The accompanying data streams are responsible to shuffle the required state of an event to the executing AC.
Next, we focus on two important key design principles that underpin the architecture-less DBMS:

\vspace{-1.5ex}
\begin{itemize*} 
\item \textbf{Fully Stateless / Active Data:}
ACs are designed to be fully stateless meaning that events can be processed in any AC and all state (e.g., table data but also any other state such as catalog data or statistics) required to execute an event is being delivered to the AC via data streams.
By making ACs fully stateless, we gain a high degree of freedom as any DBMS function can be executed anywhere. This feature allows \mimicking{} diverse architectures but also to support elasticity for all database functions individually,
i.e., additional ACs can execute any DBMS function at any time.
Moreover, in architecture-less DBMSs data is active, meaning that data is not pulled after an event is scheduled but it is pushed from data sources actively to the ACs before it is actually needed (cf. Section \ref{sec:overview:challenges}).

\item \textbf{Non-blocking / Asynchronous Execution:}
A second key aspect is that ACs are executing events in a non-blocking manner. 
This means that an AC never waits for data of an event if data is not available yet.
Instead, another event whose data is available is being processed.
For example, an operator (e.g., a filter or a join) is only processed once its input data (e.g., a batch of tuples) has arrived via the data stream. 
To provide a non-blocking execution, ACs use queues to buffer input events and data items.
In addition, queues decouple the execution between ACs as much as possible; i.e., ACs can process events asynchronously from each other.
This asynchronous execution model, which is only implicitly synchronizing the execution across ACs through events and data streams, opens up many new opportunities, as we discuss later in this paper.
\end{itemize*}
\vspace{-1.5ex}

At a first glimpse, the execution model of an architecture-less DBMS seems to have similarities with existing approaches such as scalable stream processing systems or function-as-a-service offerings (aka server-less computing).
However, there are crucial differences.
(1) First and foremost, while \sysname{} also uses streams as a major abstraction, \sysname{} is different from stream processing engines, since we target classical database workloads that process relations and streams serve as a vehicle to on-the-fly adapt the database architecture on a per-query basis.
Still, our approach benefits from techniques of scalable stream processing such as efficient data routing.
(2) Similar to function-as-a-service, \sysname{} relies on a fully stateless execution model.
However, in architecture-less DBMSs data (i.e., state) does not come as an afterthought.
In classical server-less computing, a function is scheduled first and then data must be pulled in from storage
before the execution can actually start.
Instead, as mentioned before, in architecture-less DBMSs data is active, meaning that data is actively pushed from data sources to the ACs before the event is actually being processed.
Moreover, while architecture-less DBMSs logically disaggregate the DBMS execution into small functions like function-as-a-service, we still allow executing events in a physically aggregated manner and also allow shipping events to the data to make use of locality.  

\subsection{Supporting OLAP and OLTP}
\label{sec:overview:oltpolap}

In the following, we give a brief overview of how the execution model above can be used to execute OLAP and OLTP workloads.

\begin{figure}
    \centering
    \includegraphics[width=1\linewidth]{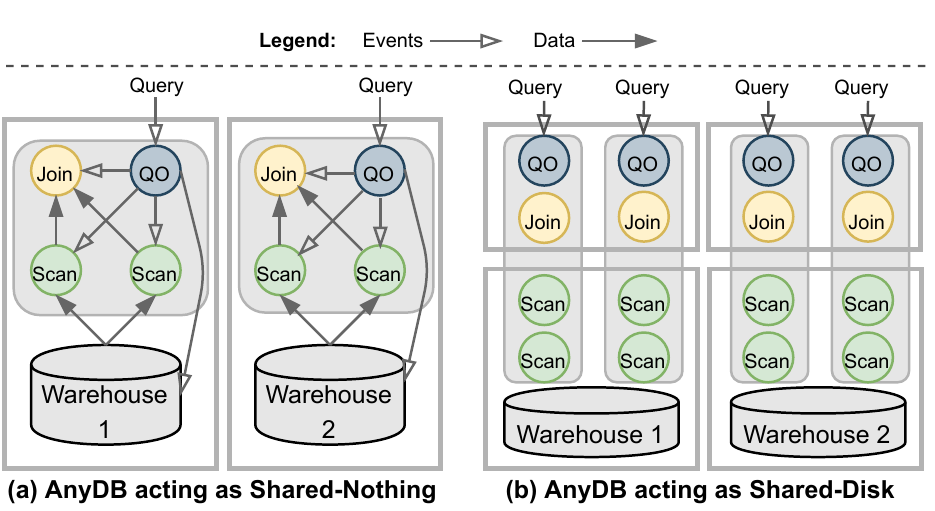}
   \vspace{-4.5ex}
   \caption[]{\sysname{} can \mimic{} diverse architectures simply by using different routing schemes for events and data streams. In (a), two servers act as a shared-nothing database while in (b) additional resources (i.e., two servers with additional 4 ACs per server) are added and \sysname{} acts as a disaggregated architecture to deal with a higher query load. For simplicity, we only show the events and data streams for (a). The gray-shaded boxes around the ACs, however, indicate in (b) which ACs execute events of the same query.}
	\label{fig:anydb2}
	\vspace{-3.5ex}
\end{figure}

\paragraph{Supporting OLAP:} The basic flow when executing an OLAP query in an architecture-less DBMS is shown in Figure \ref{fig:anydb2}.
The initial event is typically a SQL query that is sent from a client to any AC of the DBMS -- never mind which one -- which then acts as the Query Optimizer (QO).
The main task of the QO is to come up with an efficient execution plan like a traditional query optimizer in a static DBMS architecture.

In contrast to traditional optimizers, however, the QO in an architecture-less DBMS produces an event stream and initiates the data streams that instrument the ACs for query execution.
Importantly, the QO also determines the routing of a query's events and data through the architecture-less DBMS.
Consequently, by these routing decisions the QO defines the DBMS architecture perceived by individual queries.

For example, as shown in Figure \ref{fig:anydb2} (a), if a query touches only one partition and there is moderate load in the system, then the QO can route events of a query such that the architecture-less DBMS acts as a shared-nothing architecture.
However, in case the query load in the system increases, servers with additional ACs are added and the architecture-less DBMS executes queries in a disaggregated mode simply by routing events differently, as shown in Figure \ref{fig:anydb2} (b).

\paragraph{Supporting OLTP:} The basic flow of executing OLTP transactions is similar to executing OLAP read-only queries.
Transactions are also decomposed into event and data streams where routing decisions define the architecture.
A key difference to read-only OLAP, however, is that in OLTP (1) transactions need to update state and (2) concurrently running transactions need to coordinate their operations to guarantee correct isolation.
Both these aspects are discussed below in the following
(cf. \emph{Concurrency and Updates}).

\subsection{Key Challenges}
\label{sec:overview:challenges}

There are different key challenges to enable efficient execution in an architecture-less DBMS.
One of them is the optimal routing of events and data for a given workload.
Another one is to handle concurrency and updates.
In the following, we briefly discuss the main ideas how we aim to address these challenges. Some of these ideas are already built into our prototype DBMS \sysname{} while others represent future routes of research.

\paragraph{Event and Data Routing:}
As mentioned before, a key challenge of an architecture-less DBMS is to decide how to handle a query and how to route its events, as part of the query optimization.
Depending on requirements of an application (e.g., latency guarantees), load in the system, and the workload, the query optimizer has to define an optimal event routing.
In our current prototype, we did not focus on this problem but use an optimal decision to showcase the potential of our approach.
We believe however that this is an interesting avenue for learned query optimizers.

A second challenging aspect is the efficient data routing.
As mentioned before, this aspect is important for latency hiding.
We utilize the decoupling of data streams from events in our execution model to solve this challenge.
The main observation is that in DBMS execution one often knows which data is accessed way ahead of time before the data is actually being processed.
For example, complex OLAP queries need to be optimized and compiled, often taking up to 100ms in commercial query optimizers in our experience, while we already know which tables contribute to a query before query optimization.
In \sysname{} we make use of this fact and initiate data streams as early as possible.
Once initiated a data stream actively pushes data to the AC where, for example, a filter operator will be executed once query optimization finished. 
We call this feature \emph{data beaming} as data is often available at an AC before the according event arrives, entirely hiding latencies of data transfers.
We analyze the opportunities of data beaming for OLAP later in Section \ref{sec:olap}.

\paragraph{Concurrency and Updates:} In general, updates are supported in \sysname{} by event streams directed towards the storage which ingests these events and produces acknowledgment events when the updates have been processed, as required for transaction coordination in OLTP.
A major challenge in handling updates thus is to hide latencies of updates as much as possible.
Again, to hide latencies of updates and decrease the overall latency of executing a transaction, operations of one transaction are represented as events and executed asynchronously by ACs.
For example, an update can be sent to the storage by one AC while other (independent) operations of the transaction can progress on other ACs.
Only the commit operation at the end of a transaction needs to know if the write successfully persisted and thus needs to wait for the acknowledgment event coming from the storage.
As we show later in Section \ref{sec:oltp}, this asynchronous model for OLTP provides many interesting opportunities and results in higher performance under various workloads.

Another challenge that is harder to solve is to efficiently handle concurrency. 
A \naive{} way would be to implement a lock manager using events representing lock operations and data streams providing the state of the lock table.
A more clever way, however, is to rethink concurrency protocols and route events and data streams such that their processing order already captures the needs of a particular isolation level for concurrency control, as we discuss later in Section \ref{sec:oltp}.

\paragraph{Fault-Tolerance and Recovery:} Fault-tolerance and recovery are two major challenges any DBMS needs to address.
For an architecture-less DBMS this is a challenge due to the asynchronous (decoupled) execution of multiple ACs where individual ACs might fail. 

Again, a \naive{} approach would be to implement standard write-ahead logging by sending log events from ACs to durable storage.
For recovery, the DBMS could be stopped and the log could be used to bring the DBMS into a correct state.
Again, in an architecture-less DBMS we believe that we can do better and learn from the streaming community.
For example, as the entire execution of a DBMS is represented as streams, another direction is to make the streams reliable, such that upon AC failure the streams (events and data) can be rerouted to another AC.
Applying these ideas is again an interesting avenue of future research.
\section{Opportunities for OLTP}
\label{sec:oltp}

In the following, we discuss the various opportunities emerging from an architecture-less DBMS when executing OLTP workloads and show initial results when compared to existing execution models of static architectures (shared-nothing and shared-disk).
For all initial experiments in this paper, we use the two dominant transactions of the TPC-C benchmark (i.e., payment and new-order).

\begin{figure}
    \centering
    \includegraphics[width=1\linewidth]{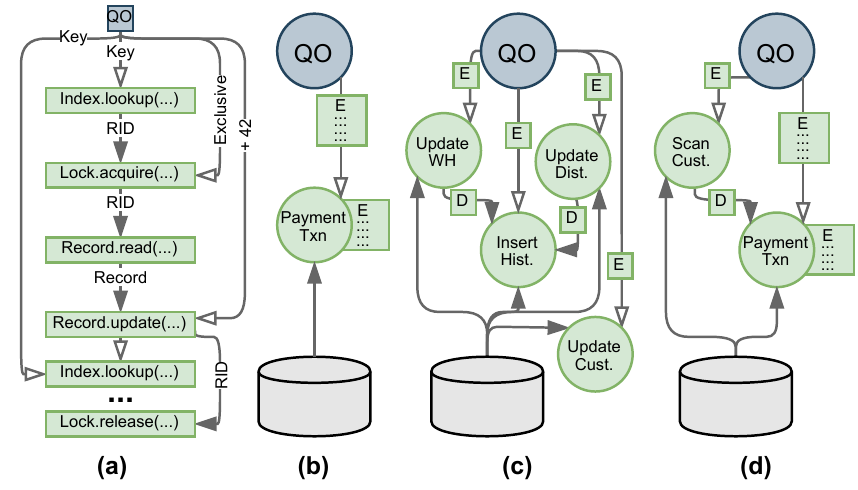}
\vspace{-4.5ex}
    \caption{Duality of Disaggregation. (a) shows how a transaction is logically disaggregated into individual events for each operation. (b) shows physically aggregated execution of events by routing the stream to one AC. (c) shows event routing for \emph{fully} intra-transaction parallel execution. (d) shows \emph{precise} intra-transaction parallel execution.}
    \Description{Duality of Disaggregation. (a) shows how a transaction is logically disaggregated into individual events for each operation. (b) shows physically aggregated execution of events by routing the stream to one AC. (c) shows event routing for \emph{fully} intra-transaction parallel execution. (d) shows \emph{precise} intra-transaction parallel execution.}
    \label{fig:oltp-architectures}
     \vspace{-3.5ex}
\end{figure}

\subsection{Opportunity 1: Duality of Disaggregation}

As indicated earlier in Figure \ref{fig:headline}, for partitionable OLTP workloads an architecture-less DBMS can achieve nearly the same throughput as an (aggregated) shared-nothing architecture.
Key to this is the duality of disaggregation in the architecture-less DBMS.
The architecture-less DBMS distinguishes logical disaggregation of the DBMS design and physical disaggregation of the DBMS execution.
Logically, the DBMS is entirely disaggregated into independent fine-grained functionality interacting via events and data streams.
However, while the overall execution is disaggregated into many small events, physically the events can still be executed in an aggregated manner if desired. This opens up the opportunity to achieve high data locality if desired as all events can be executed close to the data, e.g., a partition of the database.

For example, an OLTP transaction may consist of an event stream like in Figure \ref{fig:oltp-architectures} (a).
Yet, this logical disaggregation does not mandate disaggregated execution.
In the contrary, any sub-sequence of these events can be physically aggregated and executed by a single AC.
As shown in Figure \ref{fig:oltp-architectures} (b), the entire event stream of a transaction can be executed by a single AC.
In fact, this physical aggregation of events establishes a shared-nothing architecture that performs on par with the static shared-nothing architecture of DBx1000 as shown in Figure \ref{fig:headline}.

\textbf{The Gist:} Through this duality of disaggregation (logical vs. physical), we believe that an architecture-less DBMS can efficiently \mimic{} diverse architectures ranging from entirely aggregated shared-nothing to fine-grained disaggregated as needed, simply shifting between those by adapting event and data routes.

\subsection{Opportunity 2: Execution Strategies}

Along with the freedom of achieving different architectures on the macro-level,
the execution model in an architecture-less DBMS also provides broad freedom on the micro-level to layout parallel execution strategies in an optimal manner.

Generally, as explained earlier, transactions are represented as event streams flowing through the architecture-less DBMS.
Importantly for transaction execution,
this event-based execution allows diverging from typical execution models in OLTP that aim for \emph{inter}-transaction parallel execution and allows investigating also other forms of parallelism.
For example, event-based execution naturally brings opportunities to ad hoc parallelize execution within a single transaction to achieve \emph{intra}-transaction parallelism,
especially when contention prohibits inter-transaction parallel execution.

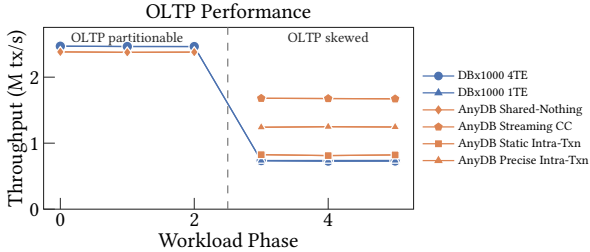
\begin{figure}
    \centering
    \begin{tikzpicture}
        \setlength\figW{6.5cm}
        \setlength\figH{4cm}
\definecolor{color0}{rgb}{0.298039215686275,0.447058823529412,0.690196078431373}
\definecolor{color1}{rgb}{0.866666666666667,0.517647058823529,0.32156862745098}

\begin{axis}[
title style = {text depth=0ex},
axis line style={white!15!black},
height=\figH,
legend cell align={left},
legend entries={DBx1000 4TE, DBx1000 1TE, AnyDB Shared-Nothing, AnyDB Streaming CC, AnyDB Static Intra-Txn, AnyDB Precise Intra-Txn},
legend style={fill=none, text opacity=1, at={(1,0.5)}, anchor=west, draw=none, nodes={scale=0.6, transform shape}},
tick pos=left,
title style={yshift=-1.5ex},
title={OLTP Performance},
width=\figW,
x grid style={white!80!black},
xlabel shift=-1ex,
xlabel={Workload Phase},
xmajorticks=true,
xmin=-0.25, xmax=5.275,
xtick style={color=white!15!black},
xticklabel style={align=left},
y grid style={white!80!black},
ylabel shift=-1.4ex,
ytick scale label code/.code={},
ylabel={Throughput (M tx/s)},
ymin=0, ymax=2753743.97436683,
yminorticks=false,
ytick style={color=white!15!black},
]
\path [draw=color0, line width=0.48pt]
(axis cs:0,2464100.2801027)
--(axis cs:0,2473585.7740863);

\path [draw=color0, line width=0.48pt]
(axis cs:1,2459398.72376511)
--(axis cs:1,2465583.06741841);

\path [draw=color0, line width=0.48pt]
(axis cs:2,2458627.03878892)
--(axis cs:2,2469107.83220649);

\path [draw=color0, line width=0.48pt]
(axis cs:3,725725.727201536)
--(axis cs:3,741116.999386462);

\path [draw=color0, line width=0.48pt]
(axis cs:4,728100.359760864)
--(axis cs:4,736059.579628886);

\path [draw=color0, line width=0.48pt]
(axis cs:5,712855.221080549)
--(axis cs:5,738640.219401011);

\path [draw=color0, line width=0.48pt]
(axis cs:3,735445.532903833)
--(axis cs:3,736684.881200355);

\path [draw=color0, line width=0.48pt]
(axis cs:4,736077.553130998)
--(axis cs:4,736812.438867593);

\path [draw=color0, line width=0.48pt]
(axis cs:5,736142.576094138)
--(axis cs:5,737449.530797736);

\path [draw=color1, line width=0.48pt]
(axis cs:0,2375032.47114707)
--(axis cs:0,2390694.81809427);

\path [draw=color1, line width=0.48pt]
(axis cs:1,2372400.09044775)
--(axis cs:1,2379745.68698982);

\path [draw=color1, line width=0.48pt]
(axis cs:2,2376528.21904461)
--(axis cs:2,2388582.87244258);

\path [draw=color1, line width=0.48pt]
(axis cs:3,1659510.02966374)
--(axis cs:3,1688661.48247592);

\path [draw=color1, line width=0.48pt]
(axis cs:4,1655567.46644372)
--(axis cs:4,1706397.28341552);

\path [draw=color1, line width=0.48pt]
(axis cs:5,1650020.36743891)
--(axis cs:5,1688107.38895155);

\path [draw=color1, line width=0.48pt]
(axis cs:3,820936.760675898)
--(axis cs:3,840595.288568482);

\path [draw=color1, line width=0.48pt]
(axis cs:4,796475.991973513)
--(axis cs:4,830464.769608311);

\path [draw=color1, line width=0.48pt]
(axis cs:5,817816.542281882)
--(axis cs:5,833370.227327772);

\path [draw=color1, line width=0.48pt]
(axis cs:3,1236348.96876905)
--(axis cs:3,1253004.2735277);

\path [draw=color1, line width=0.48pt]
(axis cs:4,1230228.87639352)
--(axis cs:4,1255381.46727417);

\path [draw=color1, line width=0.48pt]
(axis cs:5,1235622.37526778)
--(axis cs:5,1256324.81021643);

\addplot [line width=0.48pt, color0, mark=*, mark size=2, mark options={solid,draw=white}]
table {%
0 2468935.80098019
1 2464420.6923849
2 2463741.93052356
3 736083.717521005
4 728692.796478427
5 730132.873875364
};
\addplot [line width=0.48pt, color0, mark=triangle*, mark size=2, mark options={solid,rotate=0,draw=white}]
table {%
3 736333.647502356
4 736720.158689522
5 737051.389065474
};
\addplot [line width=0.48pt, color1, mark=diamond*, mark size=2, mark options={solid,draw=white}]
table {%
0 2383030.44023509
1 2376234.38878982
2 2379494.40948942
};
\addplot [line width=0.48pt, color1, mark=pentagon*, mark size=2, mark options={solid,draw=white}]
table {%
3 1679943.46990224
4 1676198.2984492
5 1672149.92496227
};
\addplot [line width=0.48pt, color1, mark=square*, mark size=1.5, mark options={solid,draw=white}]
table {%
3 825868.258527735
4 811856.762081824
5 822068.004548091
};
\addplot [line width=0.48pt, color1, mark=triangle*, mark size=2, mark options={solid,draw=white}]
table {%
3 1241196.23390033
4 1249322.63288498
5 1244921.11090535
};
\addplot [line width=0.48pt, white!50.1960784313725!black, dashed, forget plot]
table {%
2.5 0
2.5 2753743.97436683
};
\addplot [line width=0.48pt, color0, forget plot]
table {%
0 2468935.80098019
1 2464420.6923849
2 2463741.93052356
3 736083.717521005
4 728692.796478427
5 730132.873875364
};
\addplot [line width=0.48pt, color0, forget plot]
table {%
3 736333.647502356
4 736720.158689522
5 737051.389065474
};
\addplot [line width=0.48pt, color1, forget plot]
table {%
0 2383030.44023509
1 2376234.38878982
2 2379494.40948942
};
\addplot [line width=0.48pt, color1, forget plot]
table {%
3 1679943.46990224
4 1676198.2984492
5 1672149.92496227
};
\addplot [line width=0.48pt, color1, forget plot]
table {%
3 825868.258527735
4 811856.762081824
5 822068.004548091
};
\addplot [line width=0.48pt, color1, forget plot]
table {%
3 1241196.23390033
4 1249322.63288498
5 1244921.11090535
};
\draw (axis cs:1,2533444.45641748) node[
  scale=0.6,
  anchor=base,
  text=white!15!black,
  rotate=0.0,
  align=center
]{OLTP
partitionable};
\draw (axis cs:4,2533444.45641748) node[
  scale=0.6,
  anchor=base,
  text=white!15!black,
  rotate=0.0,
  align=center
]{OLTP
skewed};
\end{axis}
    \end{tikzpicture}

    \vspace{-3ex}
    \caption{OLTP performance of \sysname{} versus the shared-nothing DBx1000 under partitionable and skewed OLTP. In phases 0-2 \sysname{} acts as a shared-nothing DBMS using an inter-transaction parallel execution model while in phases 3-5 \sysname{}  acts as a shared-disk DBMS using an intra-transaction parallel execution model. Note that for DBx1000, 4 transaction executors (TEs) perform like a single TE due to high contention between transactions.}
    \label{fig:oltp-performance}
    \vspace{-2.5ex}
\end{figure}

The efficiency of this freedom to change the transaction execution on the micro-level becomes also visible in Figure \ref{fig:oltp-performance}.
When running a partitionable OLTP workload in the first phase (0-2), \sysname{} \mimics{} not only a shared-nothing architecture but also uses classical inter-transaction parallelism. Afterwards it uses intra-transaction parallelism in the second phase (3-5) for the highly skewed, contended OLTP workload,
where $100\%$ of TPC-C payment transactions operate on one warehouse only.

The baseline DBx1000 in this experiment uses a shared-nothing model and is thus bound by the resources that are assigned to one partition resulting in 0.7 M txn/s.
In such a case, our architecture-less DBMS allows to simply shift from inter- to intra-transaction parallelism by routing events of a single transaction to several ACs, i.e., shifting from Figure \ref{fig:oltp-architectures} (b) to (c).
This intra-transaction parallel execution may accelerate transactions in contended OLTP workload,
as already proven by architectures such as DORA~\cite{dora}.

However, just like any design decisions in a static architecture,
also static intra-transaction parallelization does not always prove beneficial either.
For example, Figure \ref{fig:oltp-performance} shows that with \naive{} intra-transaction parallelism, where every independent operation of a transaction is farmed out to a different AC, \sysname{} only achieves 0.8 M txn/s (orange squares), barely exceeding the inter-transaction parallel execution of the DBx1000 baselines (blue lines).
The main reason is that the overhead of parallelizing within one transaction dominates the execution.

The architecture-less DBMS addresses the challenging parallelization of transactions in the following ways:
the representation of transactions as event streams allows the architecture-less DBMS to route independent sub-sequences of events (i.e., sub-sequences of operations) to individual ACs for parallel execution.
Thus, the main challenge in an architecture-less DBMS is to split a transaction into suitable sub-sequences of events and route them to different ACs, balancing the amount of work versus overhead.
In our experiments, for example, we partition the TPC-C payment transaction into one sub-sequence with several brief update statements and a second sub-sequence with a long range scan, as depicted in Figure \ref{fig:oltp-architectures} (d).

Finding an optimal splitting and routing of event sequences of transactions depends on many factors.
One important point is that the individual sub-sequences of operations have similar execution latencies, such that the overall latency of the transaction is minimized.
Finding such an optimal routing of events is an opportunity for learning query optimization and scheduling, as mentioned before. 
As shown in Figure \ref{fig:oltp-performance}, with this balanced (i.e., precise) intra-transaction parallelization \sysname{} achieves 1.2 M txn/s (orange triangle) with only 2 ACs, outperforming the baseline (blue lines) and \sysname{}'s \naive{} parallelization (orange square) by 3.2x and 3x in throughput per thread, respectively.

\textbf{The Gist:} Generally, we envision, that the freedom of the execution models on the micro-level in an architecture-less DBMS can enable new parallel transaction execution models spanning any design between pure inter- to aggressive (fine-grained) intra-transaction parallelism on a per-transaction level.

\subsection{Opportunity 3: Concurrency Control}

In OLTP workloads, especially under high contention, concurrency control (CC) causes significant coordination effort and challenges efficient parallelization.
In an architecture-less DBMS, the event-based nature of transactions provides the opportunity to transform CC to a streaming problem.
Thereby, the architecture-less DBMS can improve the efficiency of traditional CC schemes and opens many opportunities for CC schemes, as discussed next.

\paragraph{Transforming Traditional Concurrency Control:}

Interestingly, many traditional CC protocols are stream-like already and thus benefit from a direct mapping to the asynchronous (non-blocking) execution model.
For example, a pessimistic lock-based CC scheme needs to match incoming lock requests with its lock state.
This can be mapped to a streaming join on an event stream containing lock requests and a data stream containing the lock state of the requested item.
Similarly, verification in optimistic CC protocols joins the read/write set of a transaction which is one data stream with the current state of the database which is another data stream.
Yet, an architecture-less DBMS also offers opportunities for novel coordination-free CC schemes vastly outperforming these traditional approaches, to be explained in the following.

\paragraph{Novel Streaming Concurrency Control:}

The key idea of rethinking CC schemes is that they can be enabled by efficiently ordering events of (conflicting) transactions flowing through the architecture-less DBMS,
rather than actively synchronizing execution of concurrent transaction using traditional CC schemes.
Especially under high contention, traditional CC causes high coordination overheads.
Here, the streaming execution of transactions brings new opportunities despite high contention.

Concurrency control in \sysname{} can be implicitly and coordination-free encoded into event routes.
That is, for consistency of concurrent transactions it suffices to route their events in a consistent order through ACs which execute the conflicting operations.
For example, considering two TPC-C payment transactions accessing the same warehouse,
\sysname{} can guarantee consistency by simply routing their events to all involved ACs in the same order.
Thereby, \sysname{} enables intra-transaction parallelism, routing independent events through different ACs and also provides CC without the need to actively synchronize operations at the same time.

Note that event-ordering does not violate our non-blocking (asynchronous) execution model of ACs, since operations (i.e., events) of conflicting transactions simply remain in the ACs' input queues for ordering while other events can still be executed.
In Figure \ref{fig:oltp-performance}, we see that this instantiation of \sysname{} called \emph{streaming CC} (orange pentagon), yields 1.7 M txn/s for TPC-C payment under high contention. This is much closer to the performance of the uniform (partitionable) execution of TPC-C payment in phases 0-2.

\textbf{The Gist:} Along the properties of event streams, we envision novel CC protocols, that avoid active synchronization, as discussed.
Moreover, the streaming CC enables new directions where events are gradually rerouted depending on the load, e.g., at first all events for a specific transaction are routed to a single AC until this AC becomes overloaded. Then \sysname{} may transparently repartition event streams while still keeping consistent event order.
\section{Opportunities for OLAP and HTAP}
\label{sec:olap}

Previously, we have described that the execution model of an architecture-less DBMS provides many opportunities for OLTP.
In the following, we discuss further opportunities for OLAP as well as mixed HTAP workloads. 

Especially for OLAP, operations encoded in the event streams are data intensive (e.g., a join of two large tables).
Therefore, data streams must efficiently bring data to wherever events are executed as if data access was local.
While this aspiration of ``omnipresent'' data appears challenging,
we observe that in DBMSs one often knows data to be accessed way ahead of time before actually processing it.
For example, in OLAP data is only accessed and processed after several milliseconds of query optimization.
Hence, we propose \emph{data beaming}, a technique initiating data streams early and pushing data to ACs where events will be executed.

\begin{figure}
    \centering
    \begin{tikzpicture}
        \setlength\figW{3.3cm}
        \input{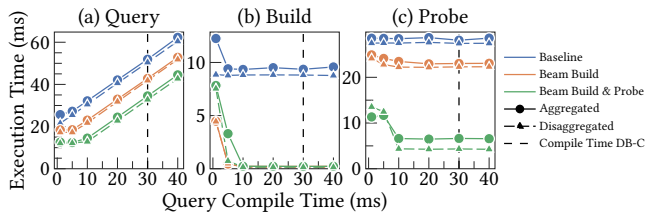}
    \end{tikzpicture}

    \vspace{-2.5ex}
    \caption{Data beaming can effectively shorten query execution for disaggregated execution of OLAP workloads and in the best case hide latencies of data shuffling completely.}
    \label{fig:olap_data_beams}
    \vspace{-2.5ex}
\end{figure}

In the following experiment, we demonstrate the effect of data beaming with a simple OLAP query.
Based on CH-benCHmark Q3~\cite{RN400}, our query reports all open orders for all customers from states beginning with ``A'' since 2007 via 3 (filtered) scans and 2 joins.
In Figure \ref{fig:olap_data_beams}, we see the effect of data beaming in several degrees for this query:
(1) In blue, the baseline does not utilize beaming, but instead passively pulls in data when needed.
(2) In orange, the build sides are beamed during query compilation, then joins are executed.
(3) In green, build and probe sides are beamed.

To detail the efficiency of data beams, we implement two variants: one where data beams only need to shuffle locally (not over the network) and one where we shuffle data across the network.
The solid lines in Figure \ref{fig:olap_data_beams} demonstrate the runtime using local beaming via shared-memory queues\footnote{\href{https://github.com/facebook/folly/blob/d2c64d94c7e892925a02a080c886ab3df3f5c937/folly/ProducerConsumerQueue.h}{Github: Folly single-producer-single-consumer queue}} (e.g., to hide NUMA latencies) and the dashed lines show the beaming across the network for a disaggregated architecture using DPI-flows \cite{cidr-2019}.
On the x-axis, 30ms marks the time taken by a commercial DBMS (DB-C) to compile this query.

Figure \ref{fig:olap_data_beams} (a) shows that the overall query execution time with beaming is only slightly higher than query compile time (green line), whereas without beaming (blue line) the query execution time has additional latency of $20$ms, since data transfer is not overlapped with query compile time.
In detail, Figures \ref{fig:olap_data_beams} (b) and (c) show the individual effects of data beaming on the build and probe side (without query compilation overhead), respectively.
We see that beaming can reduce the runtime of the smaller build side almost to $0$ms.
For the larger probe side, beaming also reduces the runtime from $30$ms to less than $10$ms.
Notably, the disaggregated architecture (which needs to shuffle data across the network) performs even better than the aggregated architecture, as DPI offloads event and data transfers to the InfiniBand NICs and acts as a co-processor in \sysname{}'s processing model. 

The previous experiment demonstrated the utility of data beaming to hide data transfer latencies  in OLAP workloads.
Besides hiding transfer latencies, data beaming can also be used to achieve other goals such as resource isolation, e.g., for HTAP workloads.
The idea is that in HTAP workloads, we can use data beams to route data intensive analytical queries to additional compute resources disaggregated from storage while latency-sensitive transactions are executed close to the data.

The HTAP workloads in Figure \ref{fig:headline} (phase 6-11) outline such scenarios, where the OLAP query of the previous experiment is executed in parallel to the OLTP workload.
Here, \sysname{} executes the OLAP query independently of the OLTP workload, only sharing storage resources, whereas the OLAP query in the static DBx1000 uses the same transaction resources for OLAP queries as for the OLTP workload.
Thereby, \sysname{} simultaneously provides higher OLTP and OLAP performance than DBx1000.

\textbf{The Gist:} In general, we envision \sysname{} to establish flexible  architectures through described data beaming as well as optimal per query/transaction event routing, opening up new paths to hybrid architectures and supporting various types of deployments.
\section{Conclusions and Future Directions}
\label{sec:directions}

In this paper, we have proposed architecture-less DBMSs, a radical new approach for scale-out distributed DBMSs.
In addition to the discussed opportunities, we see many further interesting research directions arising from the architecture-less approach:

\paragraph{Elasticity for Free:}
Flexible event routing and data beaming open up opportunities, apart from resource isolation, for transparent elasticity without additional latencies.
Considering events are self-contained and state is always beamed, 
elasticity for execution of event streams just means consistent routing of events and their state to an elastic number of ACs.
Even more, as mentioned before since all events and data are both delivered as streams to ACs, these streams could be repartitioned or rerouted to distribute load in the system adaptively.

\paragraph{Transparent Heterogeneity:}
The stateless execution model in conjunction with the opportunity for  elasticity further facilitates transparent (ad hoc) integration of heterogeneous compute resources per query, including but not limited to accelerators (e.g., FPGAs or GPUs) and programmable data planes (e.g., programmable NICs or switches).
Moreover, since events fully describe what to do and data streams deliver all required state, event execution (\emph{how to do it}) can be specialized for FPGAs, etc. without any impact nor dependencies on the rest of the DBMS.

\paragraph{Crossing Clouds and More:} Finally, data beaming is an interesting concept generally hiding data transfer cost not only within but also across data centers.
This opens up opportunities for DBMS deployments across on-premise, cloud offerings, and the edge without paying significant latencies for data transfer.
For example, an architecture-less DBMS for HTAP workload may run transactions for daily business on-premise and may ad hoc beam data to cloud resources for sporadic reporting, combining the benefits of both platforms efficiently.

\bibliographystyle{abbrv}
\bibliography{bib}

\begin{thebibliography}{1}

\bibitem{cidr-2019}
G.~Alonso et~al.
\newblock {DPI:} the data processing interface for modern networks.
\newblock In {\em {CIDR}}, 2019.

\bibitem{config}
T.~Bang et~al.
\newblock Robust performance of main memory data structures by configuration.
\newblock In {\em {SIGMOD'20}}, pages 1651--1666, 2020.

\bibitem{RN400}
R.~Cole et~al.
\newblock The mixed workload ch-benchmark.
\newblock In {\em Proceedings of the Fourth International Workshop on Testing
  Database Systems}, page Article 8. ACM, 2011.

\bibitem{RN398}
A.~K. Goel et~al.
\newblock Towards scalable real-time analytics: an architecture for scale-out
  of olxp workloads.
\newblock {\em Proc. VLDB Endow.}, 8(12):1716–1727, 2015.

\bibitem{RN399}
A.~Gupta et~al.
\newblock Amazon {Redshift} and the case for simpler data warehouses.
\newblock In {\em SIGMOD'15}, page 1917–1923. ACM, 2015.

\bibitem{dora}
I.~Pandis et~al.
\newblock Data-oriented transaction execution.
\newblock {\em Proc. {VLDB} Endow.}, 3(1), 2010.

\bibitem{repart1}
R.~Taft et~al.
\newblock E-store: Fine-grained elastic partitioning for distributed
  transaction processing.
\newblock {\em Proc. {VLDB} Endow.}, 8(3):245--256, 2014.

\bibitem{RN397}
A.~Verbitski et~al.
\newblock Amazon {Aurora}: Design considerations for high throughput
  cloud-native relational databases.
\newblock In {\em SIGMOD'17}, page 1041–1052. ACM, 2017.

\bibitem{abyss}
X.~Yu et~al.
\newblock Staring into the abyss: An evaluation of concurrency control with one
  thousand cores.
\newblock {\em Proc. VLDB Endow.}, 8(3):209–220, Nov. 2014.

\end{thebibliography}

\end{document}